\documentstyle[aaspp4]{article}

\lefthead{Case and Bhattacharya}
\righthead{New $\Sigma - D$ Relation}

\begin{document}

\title{A New $\Sigma - D$ Relation and Its Application to the Galactic 
Supernova Remnant Distribution}

\author{Gary L. Case and Dipen Bhattacharya}
\affil{Institute of Geophysics and Planetary Physics, 
University of California, Riverside, CA 92521; case@tigre.ucr.edu, 
dipen@tigre.ucr.edu}
\slugcomment{Accepted for publication in the Astrophysical Journal}

\begin{abstract}
Technological advances in radio telescopes and X-ray instruments over the
last 20 years have greatly increased the number of known supernova
remnants (SNRs) and led to a better determination of their properties. In 
particular, more SNRs now have reasonably determined distances. However, many 
of these distances were determined kinematically using old rotation curves 
(based on $R_{\sun} = 10$ kpc and $V_{\sun} = 250$ km/s).  A more modern 
rotation curve (based on $R_{\sun} = 8.5$ kpc and $V_{\sun} = 220$ km/s) is 
used to verify or recalculate the distances to these remnants.  We use a 
sample of 36 shell SNRs (37 including Cas A) with known distances to derive 
a new radio surface brightness-to-diameter ($\Sigma-D$) relation. The slopes 
derived here ($\beta = -2.64$ including Cas A, $\beta = -2.38$ without Cas A) 
are significantly flatter than those derived in previous studies.  An 
independent test of the accuracy of the $\Sigma-D$ relation was performed 
by using the extragalactic SNRs in the Large and Small Magellanic 
Clouds.  The limitations of the $\Sigma-D$ relation and the assumptions 
necessary for its use are discussed. A revised Galactic distribution of SNRs 
is presented based on the revised distances as well as those calculated from 
this $\Sigma-D$ relation.  A scaling method is employed to compensate for 
observational selection effects by computing scale factors based on individual
telescope survey sensitivities, angular resolutions and sky coverage. The
radial distribution of the surface density of shell SNRs, corrected for 
selection effects, is presented and compared to previous works.  
\end{abstract}

 
\section{Introduction}

Determining the distances to the Galactic supernova remnants (SNRs) is a
difficult but important task. The in-depth study of the remnants
has depended largely on observations in the radio regime. However, radio 
telescope surveys searching for SNRs are biased by three selection 
effects: i) overlooking SNRs due to their low surface brightness,
especially in a region of higher background, ii) failure to detect SNRs due
to their small angular size and iii) an absence of uniform 
coverage of the sky (see also Green 1991). Recent observations of SNRs have 
been performed with radio telescopes with better 
sensitivities, higher angular resolutions and more complete sky coverage
than the previous generations, as well as with space-based X-ray telescopes.
The use of these new instruments has led to the discovery of new SNRs with 
lower surface brightnesses and smaller angular sizes, resulting in a 
significant increase in the number of known SNRs.  It has also led to
a better determination of the physical and observational properties of known
remnants. However, most SNRs still do not have well determined distances.

Conventionally, SNRs are classified into three basic types: shell remnants, 
characterized by diffuse, shell-like emission with steep radio spectra;  
plerionic or filled-center remnants, containing a central source with a flat 
radio spectrum but no shell structure; and composite remnants, which 
show signs of both a central source and a diffuse shell structure. Distances 
to the SNRs can be inferred from positional coincidences with \ion{H}{1}, 
\ion{H}{2}, and molecular clouds, OB associations or pulsars or from measuring
optical velocities and proper motions.  Where there is no direct distance 
determination, estimates can be made for shell remnants by utilizing the 
radio surface brightness to diameter relationship ($\Sigma-D$) (i.\,e.
Clark \& Caswell 1976; Milne 1979; Sakhibov \& Smirnov 1982). The mean 
surface brightness at a specific radio frequency, $\Sigma_{\nu}$, is a 
distance independent parameter and, to a first approximation, is an intrinsic 
property of the SNR (Shklovsky 1960). If this relationship is given by 
\begin{equation}
\Sigma_{\nu} = AD^{\beta},
\end{equation}
the distance $d$ will be proportional to $\Sigma^{1/\beta}_{\nu} \theta^{-1}$,
or, in terms of observable quantities, 
$d \propto S_{\nu}^{1/\beta}\theta^{-(1 + 2/\beta)}$, where $S_{\nu}$ is the 
flux density at the observing frequency $\nu$ and $\theta$ is the angular 
diameter of the remnant. Composite SNRs may be characterized by a similar 
relation, but only a few have known distances, and hence a reasonable 
$\Sigma-D$ relation is difficult to calculate for them.  

There exists considerable skepticism about using the $\Sigma-D$ relation to 
obtain distance estimates (e.g. Green 1984, 1991).  Green (1984) pointed out 
two significant problems with the $\Sigma-D$ relation.  First, many of the 
independently determined distances have a degree of uncertainty.  
\ion{H}{1} absorption measurements can be difficult to interpret and often
give reliably only lower distance limits.  Associations of SNRs with other 
objects such as molecular clouds, \ion{H}{1} emission regions and OB 
associations cannot always be made with great confidence. Second, studies of 
SNRs in the Large Magellanic Cloud, which are all at approximately the same 
known distance, have shown a spread in intrinsic properties (see discussion
in Green 1984). This suggests that for a given surface brightness, there may 
be a spread in linear diameters, making a unique distance estimate to 
individual remnants difficult and uncertain.  This spread in intrinsic 
properties likely results from the variety of different environments into 
which the remnants are evolving. The density and structure of the ambient 
medium and the lingering effects of the SNR progenitor star could all 
potentially effect the remnant's evolution (Allakhverdiyev et al. 1983a, 
1983b). However, in SNR samples which are thought to be observationally 
complete, there is a clear trend for the surface brightness of SNRs to 
decrease with increasing linear diameter.  By using a large number of 
distance `calibrators' with distance measurements as reliable as possible,  
a $\Sigma-D$ relation can be constructed and distance estimates can be made 
to shell remnants for which there is no other distance information available.  
However, investigators who use distances to individual SNRs based upon the 
$\Sigma-D$ relation must be aware of the inherent uncertainties and 
assumptions of this method.

The impetus for this work arose from a need to determine the radial Galactic 
SNR distribution.  The $\Sigma-D$ relation was the only means by which 
distances could be obtained for most of the known shell SNRs.  However, it 
became apparent that the $\Sigma-D$ relations most often quoted for distance 
estimates (e.g. Clark \& Caswell 1976; Milne 1979) needed to be updated in 
light of the new information available for SNRs. We show that even with the 
uncertainties involved in estimating distances to {\em individual} SNRs, it is 
possible to use the $\Sigma-D$ relation for examining {\em ensemble} 
properties of the SNRs, such as the total number and their radial 
distribution.  In this paper, revised distances are given for all SNRs in 
Green's SNR Catalogue (Green 1996a) for which previous information is 
out-dated or new information is available.  These kinematic distances were 
recalculated when necessary using a more modern rotation curve (with 
$R_{\odot} = 8.5$ kpc and $V_{\odot} = 220$ km/s).  Other distances were 
taken from new pulsar-SNR or molecular cloud-SNR associations.  A new 
$\Sigma-D$ relation for shell type SNRs is presented using a sample of 36 
`calibrators' (37 including Cas A) and the assumptions and limitations for 
the use of the $\Sigma-D$ relation are discussed. The surface brightness 
distribution of the nearby remnants and a Monte Carlo simulation are used to 
calculate scale factors in an attempt to compensate for observational 
selection effects. These results are used to derive the radial SNR surface 
density distribution in the Galaxy.

\section{SNR Distances and the $\Sigma-D$ Relation}

The positional coincidences of Galactic SNRs with \ion{H}{1}, \ion{H}{2}, and 
molecular clouds, OB associations or pulsars may provide distance estimates 
to the remnants. However, of the 215 SNRs in Green's present catalog (August 
1996 version), only 64 have independently determined distances (38 of 160  
shell type, 5 of 9 filled center, 18 of 31 composite and 3 of 15 unknown 
type). In this analysis, any shell remnant which has an associated pulsar, 
regardless of whether or not any kind of radio plerion is observed, is 
considered a composite.  Therefore, the number of shell and composite type 
SNRs given here differs from the numbers in Green's Catalogue. Using a number 
of known diameters (from the shell remnants with established distances) one 
can derive a $\Sigma-D$ relationship for a given radio frequency. For those 
shell-type remnants which have no direct distance information, we can then 
estimate their distances using this relation. Unless stated otherwise, the 
flux densities and surface brightnesses used in this analysis are referenced 
to 1 GHz.

Previous $\Sigma-D$ relations have indicated that the power law index $\beta$ 
in Eq. 1 lies in the range $-2.8$ to $-4$. Clark \& Caswell (1976) used 20 
SNRs (14 shell, 2 filled center, 2 composite, 1 of unknown type, 1 no longer 
regarded as an SNR) as distance calibrators and found $\beta = -3 $ in the 
surface brightness range $2 \times 10^{-20} < \Sigma_{408} < 5 \times 
10^{-19}$ W m$^{-2}$ Hz$^{-1}$ (they suggested $\beta \approx -10$ for 
$\Sigma_{408} < 2 \times 10^{-20}$). This surface brightness limitation
excluded the three brightest remnants including Cas A.  They also excluded 
RCW 103, for which they had what they considered a reliable distance, but was 
far away from their fit in the $\Sigma-D$ plane. In addition, 9 of the 20 
calibrators had only lower limits on the distance, and the lower limit was 
taken to be the actual distance. Milne (1979) used the 11 SNRs from Clark \& 
Caswell with known distances, added back the four remnants that had been 
excluded and added 7 more with known distances for a total of 22 SNRs (18 
shell, 3 composite, and 1 filled center) and obtained $\beta = -3.8$. 
Lozinskaya (1981) used 21 of the 22 SNRs (excluding the Crab) in Milne's list,
with revised distances to three of them, and added 5 others who's distances 
she had determined from optical observations for a total of 26 (22 shell and 
4 composite) and found $\beta = -3.45$, roughly half way between the values 
given by Clark \& Caswell and Milne.  Sakhibov \& Smirnov (1982) used 38 SNRs 
(including those from Lozinskaya) that they classify as shell-type (but 
actually includes at least 4 remnants that we classify as composite) out of a 
larger sample of 57 calibrators of all types to derive $\beta = -3.4\pm0.5$, 
agreeing closely with Lozinskaya and, within their error, with Clark \& 
Caswell and Milne as well.  For their larger sample they obtain $\beta = 
-2.8\pm0.4$. Li \& Wheeler (1984) have derived the flattest relation for 
shell SNRs with $\beta = -2.77$. Huang \& Thaddeus (1985) attempted to use a 
more homogeneous subsample consisting of 12 shell SNRs located near large 
molecular clouds and found $\beta = -3.21$. However, all of these studies 
used old rotation curves.

\begin{deluxetable}{llcccc}
\tablecaption{Shell SNRs with known distances \label{shell}}
\tablehead{
\colhead{Catalog name    } & \colhead{Other name} & \colhead{Surface Brightness} & 
\colhead{Distance} & \colhead{Diameter} & \colhead{Ref.\tablenotemark{a}} \\
\colhead{} & \colhead{} & \colhead{(W m$^{-2}$ Hz$^{-1}$ sr$^{-1}$)} & \colhead{(kpc)} 
& \colhead{(pc)} & \colhead{} }
\startdata
G4.5+6.8\dotfill    & Kepler's SNR   & 3.2$\times 10^{-19}$ & \phn4.5  & 
   \phn\phn4 & 1\nl
G13.3--1.3\dotfill  &                & \tablenotemark{b}    & \phn3.0  & 
   \phn46 & 2 \nl
G18.8+0.3\dotfill   & Kes 67         & 2.7$\times 10^{-20}$ & \phn8.1  & 
   \phn32   & $\ast$ \nl
G31.9+0.0\dotfill   & 3C391          & 1.0$\times 10^{-19}$ & \phn7.2  & 
   \phn12   & 3,4 \nl
G33.6+0.1\dotfill   & Kes 79         & 3.3$\times 10^{-20}$ & \phn7.1  & 
   \phn21   & $\ast$ \nl
G43.3--0.2\dotfill  & W49B           & 4.8$\times 10^{-19}$ & \phn7.5  & 
   \phn\phn8 & $\ast$ \nl
G46.8--0.3\dotfill  & HC30           & 9.5$\times 10^{-21}$ & \phn6.4  & 
   \phn28   & $\ast$ \nl
G49.2--0.7\dotfill  & W51            & 2.7$\times 10^{-20}$ & \phn6.0  & 
   \phn52   & 5 \nl
G53.6--2.2\dotfill  & 3C400.2        & 1.3$\times 10^{-21}$ & \phn5.0  & 
   \phn44   & $\ast$ \nl
G54.4--0.3\dotfill  & HC40           & 2.6$\times 10^{-21}$ & \phn3.3  & 
   \phn38   & $\ast$ \nl
G74.0--8.5\dotfill  & Cygnus Loop    & 8.6$\times 10^{-22}$ & \phn0.8  & 
   \phn45   & 6 \nl
G78.2+2.1\dotfill   & $\gamma$ Cygni & 1.4$\times 10^{-20}$ & \phn1.2  & 
   \phn21   & 7,8 \nl
G84.2--0.8\dotfill  &                & 5.2$\times 10^{-21}$ & \phn4.5  & 
   \phn24   & 9 \nl
G89.0+4.7\dotfill   & HB21           & 3.1$\times 10^{-21}$ & \phn0.8  & 
   \phn24   & 10 \nl
G111.7--2.1\dotfill & Cas A          & 1.6$\times 10^{-17}$ & \phn3.4  & 
   \phn\phn5 & 11 \nl
G116.5+1.1\dotfill  &                & 3.4$\times 10^{-22}$ & \phn5.0  & 
   101      & $\ast$ \nl
G116.9+0.2\dotfill  & CTB 1          & 1.2$\times 10^{-21}$ & \phn3.1  & 
   \phn31   & 12 \nl
G119.5+10.2\dotfill & CTA 1          & 6.7$\times 10^{-22}$ & \phn1.4  & 
   \phn37   & 13 \nl 
G120.1+1.4\dotfill  & Tycho's SNR    & 1.3$\times 10^{-19}$ & \phn4.5  & 
   \phn11   & 14 \nl
G132.7+1.3\dotfill  & HB3            & 1.1$\times 10^{-21}$ & \phn2.2  & 
   \phn51   & 15,16 \nl
G156.2+5.7\dotfill  &                & 6.2$\times 10^{-23}$ & \phn3.0  & 
   \phn96   & 17,18 \nl
G160.9+2.6\dotfill  & HB9            & 9.9$\times 10^{-22}$ & \phn2.2  & 
   \phn83   & $\ast$ \nl
G166.0+4.3\dotfill  & VRO 42.05.01   & 5.5$\times 10^{-22}$ & \phn4.5  & 
   \phn57   & 19 \nl
G166.2+2.5\dotfill  & OA 184         & 2.6$\times 10^{-22}$ & \phn4.5  & 
   104      & 19 \nl
G189.1+3.0\dotfill  & IC443          & 1.2$\times 10^{-20}$ & \phn1.5  & 
   \phn20   & 20 \nl
G205.5+0.5\dotfill  & Monoceros      & 5.0$\times 10^{-22}$ & \phn1.6  & 
   102      & 21 \nl
G260.4--3.4\dotfill & Puppis A       & 6.5$\times 10^{-21}$ & \phn2.2  & 
   \phn35   & 22 \nl
G296.5+10.0\dotfill & PKS 1209-52    & 1.2$\times 10^{-21}$ & \phn1.6  &  
   \phn36   & 23,$\ast$ \nl
G304.6+0.1\dotfill  & Kes 17         & 3.3$\times 10^{-20}$ & \phn7.9  & 
   \phn18   & $\ast$ \nl
G309.8+0.0\dotfill  &                & 5.4$\times 10^{-21}$ & \phn3.6  & 
   \phn23   & 7 \nl
G315.4--2.3\dotfill & RCW 86         & 4.2$\times 10^{-21}$ & \phn2.8  & 
   \phn34   & 24 \nl
G327.6+14.6\dotfill & SN1006         & 3.2$\times 10^{-21}$ & \phn2.1  & 
   \phn18   & 25,26 \nl
G330.0+15.0\dotfill & Lupus Loop     & 1.6$\times 10^{-21}$ & \phn1.2  & 
   \phn63   & 27 \nl
\tablebreak
G332.4--0.4\dotfill & RCW 103        & 4.2$\times 10^{-20}$ & \phn3.4  & 
   \phn\phn10 & $\ast$ \nl
G348.5+0.1\dotfill  & CTB 37A        & 4.8$\times 10^{-20}$ & \phn9.0  & 
   \phn39   & $\ast$ \nl
G348.7+0.3\dotfill  & CTB 37B        & 1.4$\times 10^{-20}$ & \phn9.0  & 
   \phn45   & $\ast$ \nl
G349.7+0.2\dotfill  &                & 6.0$\times 10^{-19}$ & 13.8     & 
\phn\phn9 & $\ast$ \nl
G359.1-0.5\dotfill  &                & 3.7$\times 10^{-21}$ & \phn9.2  & 
   \phn64 & 28 \nl
\enddata
\tablenotetext{a}{Distances to remnants marked with an `$\ast$' are 
recalculated from references in Table~\ref{distances}.}
\tablenotetext{b}{G13.3-1.3 is an SNR recently identified in X-rays (by 
ROSAT), and has not yet had a radio surface brightness published.}
\noindent
\tablerefs{
(1) Bandiera 1987; (2) Seward et al. 1995; (3) Reynolds \& Moffett 1993; 
(4) Wilner et al. 1998; (5) Koo, Kim \& Seward 1995; (6) Minkowski 1958; 
(7) Huang \& Thaddeus 1985; (8) Green 1989b; (9) Feldt \& Green 1993; 
(10) Tatematsu et al. 1990; (11) Reed et al. 1995; (12) Hailey \& 
Craig 1994; (13) Pineault et al. 1993; (14) Schwarz et al. 1995; 
(15) Routledge et al. 1991; (16) Normandeau et al. 1997; (17) Reich, 
F\"{u}rst \& Arnal 1992; (18) Pfeffermann, Aschenbach \& Predehl 1991; 
(19) Landecker et al. 1989; (20) Fesen 1984; (21) Odegard 1986; (22) 
Reynoso et al. 1995; (23) Roger et al. 1988; (24) Rosado et al. 1996; 
(25) Long, Blair \& van den Bergh 1988; (26) Winkler et al. 1997; 
(27) Leahy, Nousek \& Hamilton 1991; (28) Uchida et al. 1992 }

\end{deluxetable}
\clearpage

\begin{deluxetable}{lcclc}
\tablecaption{SNRs with revised distances \label{distances}}
\tablehead{
\colhead{Name} & \colhead{Type\tablenotemark{a}} & \colhead{Distance} 
 & \colhead{Method} & \colhead{Ref.} \\
\colhead{} & \colhead{} & \colhead{(kpc)} & \colhead{} & \colhead{}}
\startdata
G5.4-1.2\dotfill    & C? & 4.6  & Pulsar coincidence    & 1 \nl
G6.4-0.1\dotfill    & C  & 3.3  & Mean optical velocity & 2 \nl
G8.7-0.1\dotfill    & C?\tablenotemark{b} & 3.9  & Pulsar coincidence &    1 \nl
G18.8+0.3\dotfill   & S  & 8.1  & \ion{H}{1} absorption & 3 \nl
G21.5-0.9\dotfill   & F  & 6.3  & \ion{H}{1} absorption & 4 \nl
G33.6+0.1\dotfill   & S  & 7.1  & \ion{H}{1} absorption & 5 \nl
G34.7-0.4\dotfill   & C?\tablenotemark{b} & 3.3  & Pulsar coincidence & 1 \nl
G43.3-0.2\dotfill   & S  & 7.5  & \ion{H}{1} absorption & 6 \nl
G46.8-0.3\dotfill   & S  & 6.4  & \ion{H}{1} absorption & 7 \nl
G53.6-2.2\dotfill   & S  & 5.0  & optical kinematics    & 8 \nl
G54.4-0.3\dotfill   & S  & 3.3  & CO association        & 9 \nl
G69.0+2.7\dotfill   & C?\tablenotemark{b}& 2.5  & Pulsar coincidence &    10 \nl
G114.3+0.3\dotfill  & C?\tablenotemark{b} & 2.5  & Pulsar coincidence &    1 \nl
G116.5+1.1\dotfill  & S  & 5.0  & \ion{H}{1} absorption & 11 \nl
G130.7+3.1\dotfill  & F  & 3.3  & \ion{H}{1} absorption & 12 \nl
G160.9+2.6\dotfill  & S  & 2.2\tablenotemark{c}  & Mean optical velocity  & 2 \nl
G180.0-1.7\dotfill  & C?\tablenotemark{b} & 1.5 & Pulsar coincidence & 10 \nl
G296.5+10.0\dotfill & S  & 1.6  & associated \ion{H}{1} shell & 13 \nl
G304.6+0.1\dotfill  & S  & 7.9  & \ion{H}{1} absorption & 3 \nl
G308.8-0.1\dotfill  & C? & 8.7  & Pulsar coincidence    & 1 \nl
G320.4-1.2\dotfill  & C  & 4.4  & Pulsar coincidence    & 1 \nl
G332.4-0.4\dotfill  & S  & 3.4  & \ion{H}{1} absorption & 3 \nl
G341.2+0.9\dotfill  & C? & 6.9  & Pulsar coincidence    & 1 \nl
G343.1-2.3\dotfill  & C? & 1.8  & Pulsar coincidence    & 1 \nl
G348.5+0.1\dotfill  & S  & 9.0  & \ion{H}{1} absorption & 3 \nl
G348.7+0.3\dotfill  & S  & 9.0  & \ion{H}{1} absorption & 3 \nl
G349.7+0.2\dotfill  & S  & 13.8 & \ion{H}{1} absorption & 3 \nl
G354.1+0.1\dotfill  & C? & 4.2  & Pulsar coincidence    & 1 \nl
\enddata
\tablenotetext{a}{S - Shell remnant; F - Filled-center remnant; C - Composite
remnant.}
\tablenotetext{b}{These SNRs are classified in Green's catalog as shell or 
unknown type.  They have been classified here as possible composite type 
because of the association of the remnant with a pulsar.}
\tablenotetext{c}{Possible association with a pulsar at 1.8 kpc.}
\noindent
\tablerefs{
(1) Frail, Goss \& Whiteoak 1994; (2) Lozinskaya 1981; (3) Caswell et al. 
1975; (4) Davelaar et al. 1986; (5) Frail \& Clifton 1989; (6) Radhakrishnan 
et al. 1972; (7) Sato 1979; (8) Rosado 1983; (9) Junkes, F\"{u}rst \& Reich 
1992; (10) Anderson et al. 1996; (11) Reich \& Braunsfurth 1981; (12) Roberts 
et al. 1993; (13) Dubner et al. 1986 }
\end{deluxetable}

We have searched the literature to find recent and accurate distances to as 
many shell remnants as possible.  Many of the remnants have more than one 
distance available.  For these remnants, we have either chosen the most 
recent measurement, or used an average of the available estimates (if the 
distance range is narrow).  Our sample of 37 shell remnants with published 
distances, which is used to derive the $\Sigma-D$ relation, is listed in 
Table~\ref{shell}.  Only the references for the distances actually used 
are given in the Table.  The surface brightnesses and angular diameters (used 
to calculate the linear diameters) are taken from Green's Catalogue.

Most of the kinematic distances given in Green's Catalogue, as well as the
distances used by the authors above, were calculated using 10 kpc as the
solar distance from the Galactic Center and a solar velocity of 250 km/s.
Some of these remnants now have more recently determined distances, where
modern rotation curves with $R_{\odot} = 8.5$ kpc and $V_{\odot} = 220$ km/s
were employed.  For those SNRs that still do not have kinematic distances
based on a modern rotation curve, the original references for the distance
information were checked in order to obtain the kinematic velocities
determined from \ion{H}{1} absorption, mean optical velocities or associated
CO or \ion{H}{1} emission.  The distances were then recalculated using the
rotation curve given by Burton \& Gordon (1978), with corrections by Fich,
Blitz \& Stark (1989), for the inner Galaxy and a flat rotation curve outside
the Solar Circle.  Saken et al. (1992) used the modern rotation curve of 
Clemens (1985), with the modification that a flat curve was used outside 5 
kpc, to derive distances to several remnants.  However, this is a 
non-standard rotation curve implementation, as fiducial rotation curves 
employ a flat curve only outside the Solar Circle.  For this reason, the 
distances derived by Saken et al. are not used in Table~\ref{shell}. In total, 
kinematic distances for 17 remnants (14 shell, 1 composite, 2 filled-center) 
were recalculated and are listed in Table~\ref{distances}.  Distances to 11 
other composite remnants were assigned the distance to their associated 
pulsars.  Those shell remnants in Table~\ref{shell} which had kinematic 
distances recalculated have an `$\ast$' in the reference column.  

For distances to the other shell 
remnants in Table~\ref{shell}, two were obtained from an association of the 
SNR with an \ion{H}{2} region and an OB association (G189.1+3.0, G205.5+0.5), 
three from associations with a molecular cloud and an OB association 
(78.2+2.1, G89.0+4.7, G309.8+0.0), one from optical kinematics and historical 
observations (G4.5+6.8), three from optical proper motions (G74.0-8.5, 
G111.7-2.1, G327.6+14.6), one from the modeling of its X-ray emission 
(G330.0+15.0), one from modelling of its X-ray emission and from associated 
\ion{H}{1} emission (G119.5+10.2), two from X-ray absorption and associated 
\ion{H}{1} emission (G296.5+10.0, G156.2+5.7) and eleven from the literature 
in which a modern rotation curve had already been used (G31.9+0.0, G49.2-0.7, 
G84.2-0.8, G116.9+0.2, G120.1+1.4, G132.7+1.3, G166.0+4.3, G166.2+2.5, 
G260.4-3.4, G315.4-2.3, G351.9-0.5). Note that the distances listed in 
Table~\ref{shell} are the revised distances (where relevant). 

\begin{figure}[tb]
\begin{center}
\epsfxsize=3.5in
\epsfbox{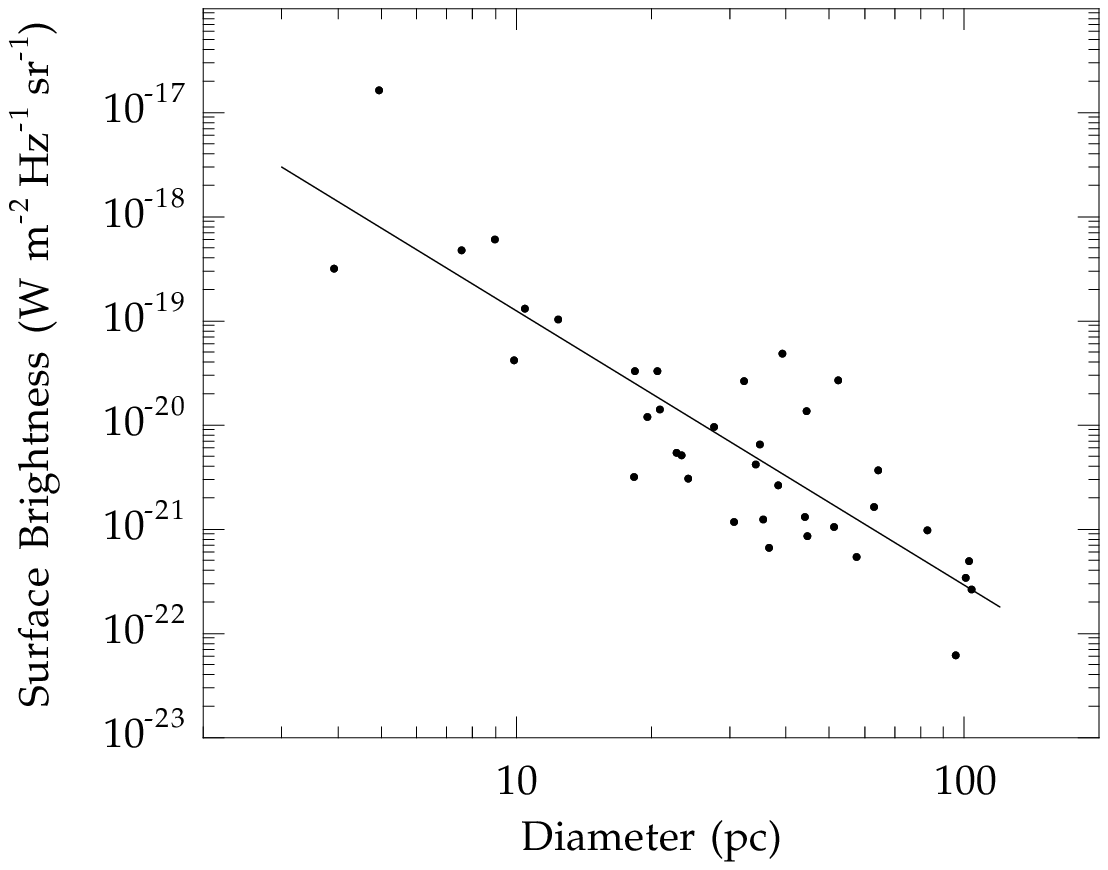}
\caption{The surface brightness versus diameter ($\Sigma - D$) 
relation for shell SNRs using the distance calibrators in Table~\ref{shell} 
(including Cas A). \label{sigma-d}}
\end{center}
\end{figure}
The diameters derived from the new distances are shown in 
Figure~\ref{sigma-d}, from which the relation
\begin{equation}
\Sigma_{{\rm 1 GHz}} = 5.43^{+8.16}_{-3.26} \times 10^{-17} 
D^{(-2.64\pm 0.26)} \mbox{ W m$^{-2}$ Hz$^{-1}$ sr$^{-1}$.}
\end{equation} 
is obtained. The typical errors in the kinematic distances are about 10-25\%, 
depending on the resolution and sensitivity of the measurements and the error 
in the rotation curve parameters.  This error 
does not include uncertainty due to the presence of noncircular motions 
associated with the object (i.e. CO cloud) or feature (i.e. \ion{H}{1} 
emission), which is hard to quantify.  The error in distances derived 
from the modeling of x-ray emission is typically 30\% (Kassim et al. 1994).
For distances estimated by associating the remnant with an object such as an 
OB association, CO cloud or \ion{H}{2} region, the distance to the object may 
be fairly well determined, but the association itself may be less certain, 
making an error estimate difficult.  In these cases, we do not incorporate the 
SNR into Table~\ref{shell} unless there is at least one other corroborating 
distance estimate.  Since it is difficult to assign realistic errors to the 
diameters, equal weighting was used in these $\Sigma-D$ fits. 

This $\Sigma-D$ relation is flatter than those obtained in the studies 
mentioned above. Some used fewer remnants than our study, some used non-shell 
remnants, and they all used old rotation curves. Radio and X-ray observations 
made in recent years have had better sensitivities and angular resolutions, 
which has led to the addition of new shell remnants with lower $\Sigma$ (with 
large $D$) and smaller $\theta$ to the current catalogue. Newly determined 
distances and revised distances using more modern rotation curves have also 
increased the number of shell remnants that can now be included in the 
$\Sigma-D$ relation.  The outlier point in Fig.~\ref{sigma-d} is Cas A.  We 
treat this remnant as a special case (see below) and derive a separate 
$\Sigma-D$ relation without Cas A.  If Cas A is excluded, we obtain the 
relation
\begin{equation}
\Sigma_{{\rm 1 GHz}} = 2.07^{+3.10}_{-1.24} \times 10^{-17} 
D^{(-2.38\pm 0.26)} \mbox{ W m$^{-2}$ Hz$^{-1}$ sr$^{-1}$}
\end{equation} 
It is this relation which will be used throughout the remainder of this paper.
When the 17 composite remnants with known distances and radio surface 
brightnesses are added, the relation becomes slightly flatter, with 
$\beta = -2.21 \pm 0.24$.

Matthewson et al. (1983a) classified SNRs based on their optical features into 
four categories:  oxygen-rich, Balmer-dominated, plerionic-composite and 
evolved. The emission from the oxygen-rich SNRs arises from shocks interacting
more with circumstellar material lost by the progenitor star in the last 
stages of evolution than with matter swept up from the interstellar medium 
(ISM), leading van den Bergh (1988a) to suggest that they arose from Type Ib 
supernova explosions, e.g. the explosion of a massive ($\gtrsim18\, 
M_{\odot}$) O or Wolf-Rayet star.  Van den Bergh (1988a) categorized the 
Galactic SNRs Cas A, Puppis A and G292.0+1.8 as oxygen-rich SNRs. Among these 
three, G292.0+1.8 is classified as a composite (Green 1996a) and Puppis A is 
an older SNR ($\sim7000$ yrs; van den Bergh 1988a).  Cas A is the only young 
oxygen rich SNR known in our Galaxy.  Its radio surface brightness is 
$\approx25$ times greater than the next brightest SNR in Table~\ref{shell}.  
Even the two shell SNRs with ages comparable to Cas A, the remnants of 
Kepler's and Tycho's supernovae, are $\approx30$ and $\approx100$ times 
fainter, respectively.  However, Kepler and Tycho, along with SN1006, are 
categorized as Balmer dominated remnants (van den Bergh 1988a).  
The Balmer-dominated remnants have filamentary shells that are strong in the 
Balmer lines of hydrogen but weak in \ion{O}{3} and \ion{S}{2} and are powered
by a high velocity, nonradiative, collisionless shock encountering the gas of 
the surrounding ISM.  It has been speculated that these remnants arise from 
Type Ia supernova explosions.  Cas A is observed to have a jet-like feature 
(Kamper \& van den Bergh 1976; Fesen, Becker \& Blair 1987), something which 
has not been observed in other shell SNRs.  The scatter in the $\Sigma-D$ 
relation (see Fig.~\ref{sigma-d}) is such that  the surface brightness of 
Cas A lies $3\sigma$ off the fit given in Eq. 2.  This suggests that young, 
Cas A-like remnants of Type Ib supernova explosions ought to be treated 
differently than the other shell SNRs.   Puppis A was retained in the fit 
as it is an older, evolved remnant.  If Puppis A is excluded, the slope of 
the $\Sigma-D$ relation does not change significantly (from $\beta = -2.379$ 
to $\beta = -2.382$).  

\begin{figure}[tb]
\begin{center}
\epsfxsize=3.5in
\epsfbox{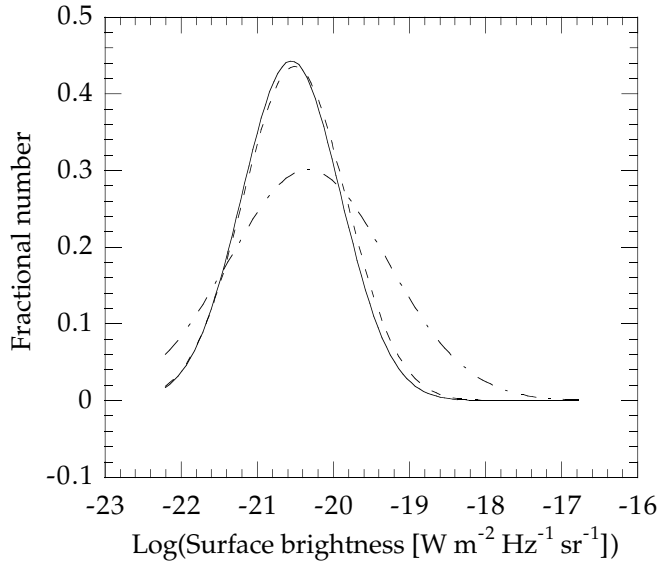}
\caption{The surface brightness distributions for all shell SNRs 
in Green's Catalog (solid line), the 20 shell SNRs within 3 kpc of the Sun 
(dashed line) and the sample of 36 shell SNRs (excluding Cas A) with known 
distances (dot dashed line).  The sample with known distances is biased 
towards brighter remnants.  The distributions are fit to gaussians with a 
lower cutoff at $\Sigma_{c} = 5 \times 10^{-23}$ W m$^{-2}$ Hz$^{-1}$ 
sr$^{-1}$.  The distribution for shell SNRs within 3 kpc of the Sun is used 
to determine scale factors to compensate for observational selection effects 
(see \S 4). \label{sb_dist}}
\end{center}
\end{figure}
The issue of selection effects in the sample used to obtain the $\Sigma-D$ 
relation must be addressed.  Figure~\ref{sb_dist} shows the fractional number 
of all shell SNRs as a function of surface brightness (solid line), along 
with the fractional number of shell SNRs used in obtaining the $\Sigma-D$ 
relation in this work (dot dashed line).  Also shown is the surface brightness
distribution for shell SNRs within 3 kpc of the Sun (dashed line). The latter 
will be used when the selection effects of the entire sample are discussed 
(see \S 4).  Apparently, the sample of shell remnants with known distances 
(i.e. Table~\ref{shell}) is underrepresented with respect to low surface 
brightness remnants as compared to the total sample.  This can be attributed 
to the fact that it is easier to determine distances to brighter remnants, 
whether because they are younger, allowing distances to be obtained from 
proper motions, or because their higher surface brightnesses allow a clearer 
interpretation of emission/absorption spectra.  

An attempt was made to compensate for this bias by utilizing a weighting 
function in the fit for the $\Sigma-D$ relation.  A weighting function was 
derived by dividing the $\Sigma$-distribution for the complete sample by the 
$\Sigma$-distribution for the SNRs with known distances.  This function then 
would allow the SNRs with lower surface brightness to have more weight in the 
fit.  The slope obtained from the weighted fit is $\beta = -1.68\pm0.65$. The 
weighted fit is not well constrained since a large 
part of the sample is effectively excluded due to the large errors assigned 
to the overrepresented part of the $\Sigma$-distribution. This has prompted 
us to use the unweighted fit of Eq. 3 throughout the remainder of this paper. 
However, it should be noted that as more fainter shell remnants have distances 
determined, the slope of the $\Sigma-D$ relation may become flatter.

\begin{figure}[tb]
\begin{center}
\epsfxsize=3.5in
\epsfbox{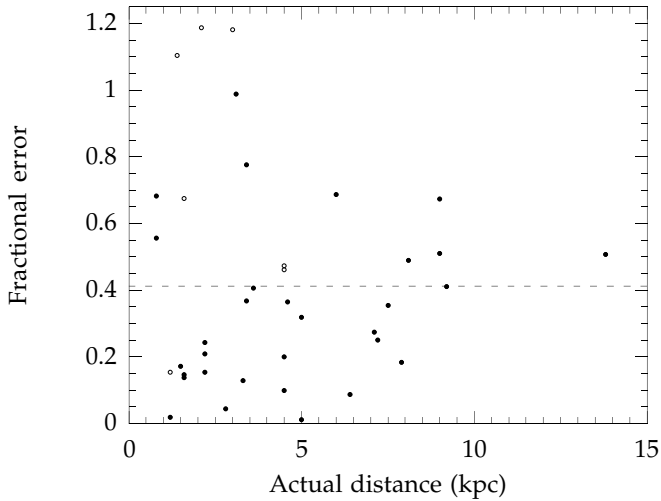}
\caption{The fractional error, $f$, of the distances derived from 
our $\Sigma - D$ relation with respect to the actual distances for the shell 
SNRs given in Table~\ref{shell}, excluding Cas A.  The open circles are those 
shell remnants with $z > 200$ pc.  The average fractional error for all of 
the remnants is 0.41 and is shown by the dashed line.  \label{frac_err}}
\end{center}
\end{figure}
In order to get an estimate of the accuracy of the $\Sigma-D$ relation for 
individual SNR distances, a fractional error was defined as 
\begin{equation} 
f = \left| \frac{d_{obs} - d_{sd}}{d_{obs}} \right|,
\end{equation}
where $d_{obs}$ is the observationally determined distance and $d_{sd}$ is 
the distance derived from the $\Sigma-D$ relation.  The fractional errors for 
the SNRs used to obtain the $\Sigma-D$ relation are plotted versus the actual 
distances in Figure~\ref{frac_err}.  The average fractional error is 
$\bar{f}=0.41$ and is represented by the dashed line in Fig.~\ref{frac_err}.  
The fractional errors are all less than $f=0.7$ with the exception of SN1006, 
CTA 1 and G156.2+5.7 which sit far off the Galactic plane ($z \approx 600$ 
pc, $z \approx 250$ pc and $z \approx 300$ pc, respectively), and CTB 1.  SNRs 
that are evolving at high $z$ are likely expanding into regions of lower 
ambient density, which could effect their surface brightness evolution.  If 
the seven SNRs with $z > 200$ pc (shown as open circles in 
Fig.~\ref{frac_err}) are excluded, the average fractional error is 
$\bar{f}=0.33$. 

An independent test of the accuracy of the $\Sigma-D$ relation was performed 
by using the extragalactic SNRs in the Large (LMC) and Small (SMC) Magellanic 
Clouds.  Matthewson et al. (1983b) presented a catalog of 25 SNRs in the LMC 
(including 3 composites) and 6 in the SMC.  They excluded the 4 
Balmer-dominated remnants and derived a $\Sigma-D$ relation for the remaining 
27 SNRs, obtaining a slope of $\beta=-2.6$.  This slope is similar to that 
derived in Eq. 3, suggesting that the Galactic and extragalactic samples 
could be compared.  Berkhuijsen (1986) gave an updated catalog of 41 shell 
SNRs in the LMC and SMC with known diameters and surface brightnesses.  
Eq. 3 was then applied to this sample and the fractional error calculated as 
above.  The average fractional error for the LMC+SMC sample was 
$\bar{f}=0.35$, less than that of the Galactic SNRs.  All of the individual 
fractional errors for the LMC+SMC sample are less than $f=0.8$ except for two 
(0505-679 and 0104-723), 
which had very large fractional errors ($f=1.7$ and $f=3.0$).  If these two 
are excluded, the average fractional error is only $\bar{f}=0.25$.

\begin{figure}[tb]
\begin{center}
\epsfxsize=3.5in
\epsfbox{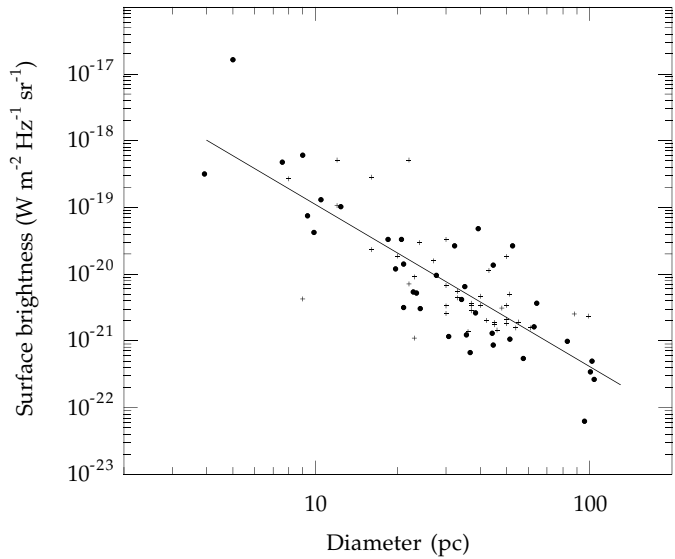}
\caption{$\Sigma - D$ relation (Eq. 5) for the Galactic (solid 
dots) and LMC and SMC (crosses) shell SNRs.  This fit includes 78 remnants 
and does not differ significantly from that with the Galactic SNRs alone 
(Eq. 3). \label{LMC}}
\end{center}
\end{figure}
A $\Sigma-D$ relation can be derived for Berkhuijsen's updated catalog. One 
SNR, 0505-679, lay far off the fit and was excluded in order to prevent it 
from unduly biasing the fit.  The slope for the remaining 40 shell SNRs is 
$\beta = -2.44\pm0.34$, very near the slope in Eq. 3.  The slope with all 41 
shell SNRs is $\beta=-2.06\pm0.35$.  The fact that i) the derived slopes of 
the $\Sigma-D$ relations for the two independent samples (Galactic and 
LMC+SMC) are consistent and ii) the application of the Galactic $\Sigma-D$ 
relation to the LMC+SMC sample yields an average fractional error of 0.25 
gives us confidence in our derived Galactic $\Sigma-D$ relation.  If the LMC 
and SMC SNRs are added to our 
sample of Galactic SNRs in Table~\ref{shell}, yielding a total sample of 78 
shell SNRs including Cas A and 0505-679, the combined fit for the $\Sigma-D$ 
relation is
\begin{equation}
\Sigma_{{\rm 1 GHz}} = 2.82^{+2.92}_{-1.43} \times 10^{-17} 
D^{(-2.41\pm 0.20)} \mbox{ W m$^{-2}$ Hz$^{-1}$ sr$^{-1}$}
\end{equation} 
and is shown in Figure~\ref{LMC}.  This is almost identical to the fit given 
in Eq. 3 but with smaller errors.
\begin{figure}[tb]
\begin{center}
\epsfxsize=3.5in
\epsfbox{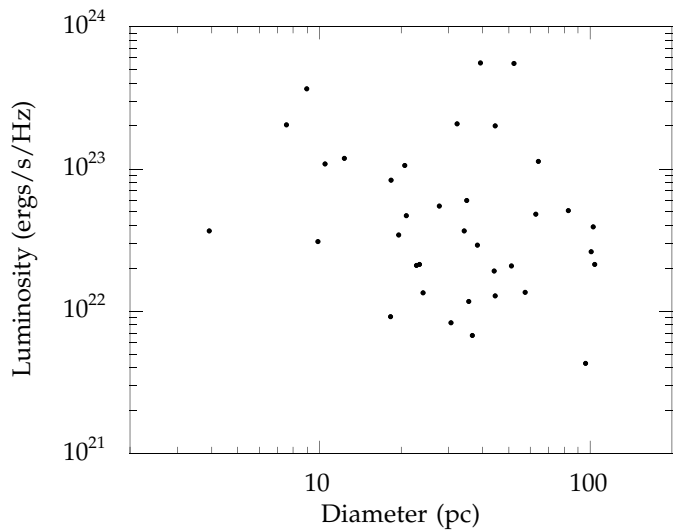}
\caption{The radio luminosity (at 1 GHz) versus the 
diameter for the shell SNRs given in Table~\ref{shell}, excluding Cas A. No 
significant correlation is evident. \label{lum_d}}
\end{center}
\end{figure}

\section{Discussion of $\Sigma-D$ Slope}

The slope, $\beta$, of the $\Sigma-D$ relation derived here is much closer  
to $\beta = -2$ than for previous relations. The radio surface brightness is 
defined as 
\begin{equation}
\Sigma_{\nu} \equiv 1.505 \times 10^{-19} \frac{S_{\nu}}{\theta^{2}} 
\mbox{ W m$^{-2}$ Hz$^{-1}$ sr$^{-1}$}
\end{equation}
where $S_{\nu}$ is in Janskys and $\theta$ is in arcminutes.  Using the 
definitions $D \propto \theta d$ ($d$ being the distance to the remnant in 
kiloparsecs) and $L_{\nu} \propto S_{\nu} d^{2}$ ($L_{\nu}$ being the radio 
luminosity per unit frequency of the remnant), Eq. 6 becomes 
\begin{equation}
\Sigma_{\nu} \propto L_{\nu} D^{-2}.
\end{equation}
If $L$ is independent of $D$, then $\beta$ must be equal to $-2$.  The 
radio luminosities of the 36 SNRs in Table~\ref{shell} (excluding Cas A) are 
plotted versus their diameters in Figure~\ref{lum_d}. Indeed, we do not find 
any significant correlation between luminosity and linear diameter for shell 
type alone or for shell plus composite type distance calibrators (see also 
Green 1991). However, the $\Sigma-D$ relation can be written as 
\begin{equation}
\Sigma_{\nu} = A D^{-2 + \delta}
\end{equation}
to allow for a possible dependence of the luminosity on the linear diameter.

Duric \& Seaquist (1986) derived a theoretical $\Sigma-D$ relation using the 
Sedov model for the evolution of the SNR blast wave and models for the 
magnetic field and particle acceleration in the remnant.  They found that the 
magnitude of $\Sigma$ depends on the initial supernova 
explosion energy, the ambient density and the strength and evolution of the 
magnetic field.  Their diameter dependence of the surface brightness is
\begin{equation}
\Sigma \propto D^{-(\gamma + \gamma \alpha + 3 \alpha - 1)}
\end{equation}
where $\gamma$ describes the magnetic field evolution and $\alpha$ is the 
radio spectral index.  For reasonable values of $\gamma$ used by Duric \& 
Seaquist ($1.5 \leq \gamma \leq 2$) and typical values for $\alpha$ 
($0.3 < \alpha < 0.8$), the slope of the $\Sigma-D$ relation 
lies in the range $-1.9 < \beta < -5$.  All of the slopes 
derived thus far fall within this range.  $\Sigma$ could also depend on the 
evolutionary stage of the shell, in which case $\delta$ would be a function 
of the age of the SNR, or on any interactions of the shell with nearby 
objects (such as molecular clouds or \ion{H}{2} regions).  Any $\Sigma-D$ 
type relation must explicitly take the factors discussed above into account 
before more accurate distances can be estimated for {\em individual} SNRs.

Several studies have been done to try and improve the $\Sigma-D$ fit.  
Caswell \& Lerche (1979) and Milne (1979) investigated the dependence of the 
radio surface brightness on the Galactic height, $z$, using separate methods. 
However, these studies rely upon only a few high $z$ remnants and therefore 
the results are not well constrained.  As noted earlier, when the SNRs with 
$z > 200$ pc were excluded, the average fractional error decreased, possibly 
indicating a different surface brightness evolution for these high $z$ 
remnants.  But when we examined the $z$-dependence of the surface brightness 
for the remnants in Table~\ref{shell}, no clear correlation was found (see 
also Green 1984, Allakhverdiyev et al. 1983b). This does not preclude a 
$z$-dependence, but more high $z$ remnants are needed to adequately address 
this question. 

Huang \& Thaddeus (1985) 
suggested using a subsample of SNRs that are evolving into similar 
environments to reduce the scatter in the $\Sigma-D$ relation.  They used 12 
SNRs (including Cas A and W44, which we consider a composite) situated near 
molecular cloud complexes and determined distances to the remnants by using 
the optical distances to \ion{H}{2} regions or OB associations associated 
with the cloud or from CO emission from the cloud itself.  The $\Sigma-D$ 
relation they derive has a slope of $-3.21\pm0.32$ with less scatter than 
previous relations.  More up-to-date distances (kinematic distances 
recalculated using a modern rotation curve or more reliable distance 
estimates) are available for some of the SNRs, and we find that the slope 
flattens to $\beta = -3.02\pm0.44$ for the same 12 remnants when the new 
distances are used (although the slopes are consistent within errors).  Huang 
and Thaddeus excluded one remnant (HB 9) simply because it lay too far off 
their $\Sigma-D$ fit.  If we use a more recent distance estimate to HB 9 (see 
Table~\ref{distances}) and then include it in the fit, we obtain 
$\beta = -2.88\pm0.40$ for the 13 SNRs.  Excluding the composite remnant W44, 
leaving 12 shell SNRs, does not significantly change the fit parameters.  
This value for the slope agrees, within the errors, to the value derived in 
Eq. 2.  Notice that using more up-to-date distances has the effect of 
flattening the slope of the $\Sigma-D$ relation.

Also, we have shown that adding the shell SNRs in the LMC and SMC to the 
Galactic sample decreases the errors in the fit for the $\Sigma-D$ relation.

\begin{figure}[tb]
\begin{center}
\epsfxsize=5.0in
\epsfbox{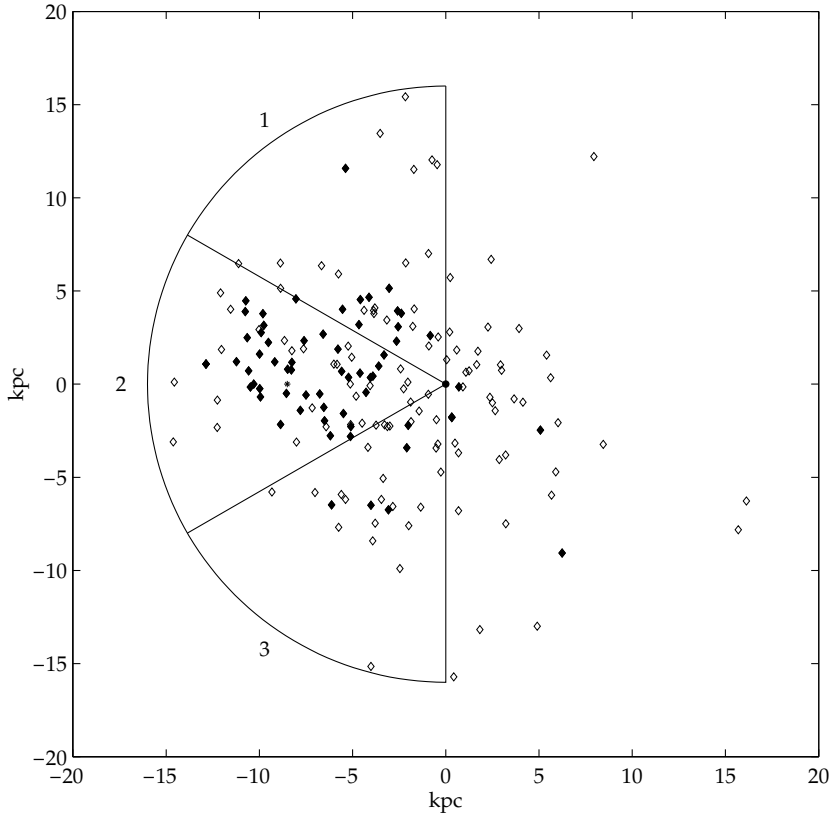}
\caption{The Galactic SNR distribution using corrected distances 
and our new $\Sigma - D$ relation. Filled-center and composite SNRs with
known distances are included. Shell SNRs with distances derived from our 
$\Sigma - D$ relation (c.f. Table~\ref{others}) are shown as open diamonds, 
while SNRs with known distances are shown as filled diamonds. The filled 
circle marks the Galactic Center and the `$\ast$' marks the position of the 
Sun. Also shown are the three regions into which the near half of the Galaxy 
is divided.  (see \S 4). \label{distrib}}
\end{center}
\end{figure}
As Eqs. 7, 8 and 9 suggest, the $\Sigma-D$ relation is {\em only} valid for 
use in estimating distances to individual SNRs when the following assumptions 
are made, namely that all shell SNR radio luminosities have the same diameter 
dependence and that all shell SNRs are born under the same initial conditions 
(same supernova explosion energy, density and structure of environment, 
magnetic field, etc.).  For the SNRs with known distances, there can be 
almost 2 orders of magnitude variation in luminosity for a given diameter 
(see Fig.~\ref{lum_d}).  It should be understood then that this variation in 
individual SNR luminosities gives rise to the uncertainty in the distance 
estimates derived from the $\Sigma-D$ relation.  However, while on average 
the errors in the individual distances are $\approx40\%$, the errors in 
ensemble studies of SNRs utilizing the $\Sigma-D$ relation can be much 
smaller (as will be shown).  If the aforementioned assumptions are made, 
keeping in mind the implied limitations, $\Sigma-D$ distances can be used to 
estimate ensemble properties such as the SNR surface density distribution and 
the total number of SNRs in the Galaxy.  Using the the $\Sigma-D$ relation in 
Eq. 3, the distances were calculated to all shell SNRs for which no other 
distance information is available.  These remnants and their distances are 
listed in Table~\ref{others}.  For reasons stated above, caution should be 
exercised in using the distances to individual SNRs in given 
Table~\ref{others}.  Figure~\ref{distrib} shows the positions of all remnants 
of all types with known (including recalculated) distances (plotted as filled 
diamonds) as well as shell SNRs with distances obtained from the new 
$\Sigma-D$ relation (plotted as open diamonds).  In total 178 remnants are 
shown. 

\begin{deluxetable}{lccc}
\tablecaption{Distances to shell SNRs calculated from $\Sigma-D$ relation
\label{others}}
\tablehead{
\colhead{Name} & \colhead{$\Sigma_{1 {\rm GHz}}$\tablenotemark{a}}          & 
\colhead{$\theta$\tablenotemark{b}} & \colhead{$d$} \\
\colhead{}     & \colhead{(W m$^{-2}$ Hz$^{-1}$ s$^{-1}$)} & 
\colhead{($\arcmin$)}               & \colhead{(kpc)}}
\startdata
G0.0+0.0\dotfill   & 1.7$\times 10^{-18}$ & 3.5$\times$2.5 & \phn3.3  \nl
G1.0-0.1\dotfill   & 4.6$\times 10^{-20}$ & 8              & \phn5.6  \nl
G1.4-0.1\dotfill   & 3.0$\times 10^{-21}$ & 10             & 14.1     \nl
G1.9+0.3\dotfill   & 6.3$\times 10^{-20}$ & 1.2            & 32.8     \nl
G3.7-0.2\dotfill   & 2.9$\times 10^{-21}$ & 11$\times$14   & 11.5     \nl
G3.8+0.3\dotfill   & 1.9$\times 10^{-21}$ & 18             & \phn9.6  \nl
G4.2-3.5\dotfill   & 6.1$\times 10^{-22}$ & 28             & \phn9.8  \nl
G5.2-2.6\dotfill   & 1.2$\times 10^{-21}$ & 18             & 11.5     \nl
G5.9+3.1\dotfill   & 1.2$\times 10^{-21}$ & 20             & 10.2     \nl
G6.4+4.0\dotfill   & 2.0$\times 10^{-22}$ & 31             & 14.1     \nl
G7.7-3.7\dotfill   & 3.4$\times 10^{-21}$ & 22             & \phn6.1  \nl
G8.7-5.0\dotfill   & 9.8$\times 10^{-22}$ & 26             & \phn8.7  \nl
G9.8+0.6\dotfill   & 4.1$\times 10^{-21}$ & 12             & 10.4     \nl
G11.4-0.1\dotfill  & 1.4$\times 10^{-20}$ & 8              & \phn9.2  \nl
G13.5+0.2\dotfill  & 2.6$\times 10^{-20}$ & 5$\times$4     & 12.7     \nl
G15.1-1.6\dotfill  & 1.1$\times 10^{-21}$ & 30$\times$24   & \phn7.9  \nl
G15.9+0.2\dotfill  & 1.9$\times 10^{-20}$ & 7$\times$5     & 11.1     \nl
G17.4-2.3\dotfill  & 1.3$\times 10^{-21}$ & 24             & \phn8.5  \nl
G17.8-2.6\dotfill  & 1.0$\times 10^{-21}$ & 24             & \phn9.2  \nl
G21.8-0.6\dotfill  & 2.6$\times 10^{-20}$ & 20             & \phn2.9  \nl
G22.7-0.2\dotfill  & 7.3$\times 10^{-21}$ & 26             & \phn3.7  \nl
G23.3-0.3\dotfill  & 1.4$\times 10^{-20}$ & 27             & \phn2.7  \nl
G24.7-0.6\dotfill  & 5.4$\times 10^{-21}$ & 15             & \phn7.4  \nl
G28.8+1.5\dotfill  & \tablenotemark{c}    & 100            & \tablenotemark{c} \nl
G30.7+1.0\dotfill  & 2.1$\times 10^{-21}$ & 24$\times$18   & \phn7.9  \nl
G31.5-0.6\dotfill  & 9.3$\times 10^{-22}$ & 18             & 12.9     \nl
G32.0-4.9\dotfill  & 9.2$\times 10^{-22}$ & 60             & \phn3.9  \nl
G32.8-0.1\dotfill  & 5.7$\times 10^{-21}$ & 17             & \phn6.3  \nl
G33.2-0.6\dotfill  & 1.6$\times 10^{-21}$ & 18             & 10.2     \nl
G36.6-0.7\dotfill  & \tablenotemark{c}    & 25             & \tablenotemark{c} \nl
G36.6+2.6\dotfill  & 4.8$\times 10^{-22}$ & 17$\times$13   & 20.6     \nl
G39.2-0.3\dotfill  & 5.6$\times 10^{-20}$ & 8$\times$6     & \phn5.9  \nl
G40.5-0.5\dotfill  & 3.4$\times 10^{-21}$ & 22             & \phn6.1  \nl
G41.1-0.3\dotfill  & 2.9$\times 10^{-19}$ & 4.5$\times$2.5 & \phn6.1  \nl
G42.8+0.6\dotfill  & 7.8$\times 10^{-22}$ & 24             & 10.4     \nl
G43.9+1.6\dotfill  & 3.6$\times 10^{-22}$ & 60             & \phn5.7  \nl
G45.7-0.4\dotfill  & 1.3$\times 10^{-21}$ & 22             & \phn9.1  \nl
G55.7+3.4\dotfill  & 4.0$\times 10^{-22}$ & 23             & 14.4     \nl
G57.2+0.8\dotfill  & 1.9$\times 10^{-21}$ & 12             & 14.3     \nl
\tablebreak
G59.5+0.1\dotfill  & 1.8$\times 10^{-20}$ & 5              & 13.3     \nl
G65.1+0.6\dotfill  & 2.0$\times 10^{-22}$ & 90$\times$50   & \phn6.6  \nl
G65.3+5.7\dotfill  & 1.1$\times 10^{-22}$ & 310$\times$240 & \phn2.1  \nl
G67.7+1.8\dotfill  & 2.6$\times 10^{-21}$ & 9              & 16.7     \nl
G69.7+1.0\dotfill  & 9.4$\times 10^{-22}$ & 16             & 14.4     \nl
G73.9+0.9\dotfill  & 2.8$\times 10^{-21}$ & 22             & \phn6.6  \nl
G82.2+5.3\dotfill  & 2.9$\times 10^{-21}$ & 95$\times$65   & \phn1.8  \nl
G84.9+0.5\dotfill  & 3.3$\times 10^{-21}$ & 6              & 22.5     \nl
G93.3+6.9\dotfill  & 2.5$\times 10^{-21}$ & 27$\times$20   & \phn6.6  \nl
G93.7-0.2\dotfill  & 1.5$\times 10^{-21}$ & 80             & \phn2.3  \nl
G94.0+1.0\dotfill  & 3.0$\times 10^{-21}$ & 30$\times$25   & \phn5.2  \nl
G112.0+1.2\dotfill & 1.2$\times 10^{-21}$ & 30             & \phn7.0  \nl
G117.4+5.0\dotfill & 9.4$\times 10^{-22}$ & 60$\times$80   & \phn3.3  \nl
G126.2+1.6\dotfill & 2.1$\times 10^{-22}$ & 70             & \phn6.1  \nl
G127.1+0.5\dotfill & 9.7$\times 10^{-22}$ & 45             & \phn5.1  \nl
G152.2-1.2\dotfill & 2.0$\times 10^{-22}$ & 110            & \phn6.6  \nl
G179.0+2.6\dotfill & 2.1$\times 10^{-22}$ & 70             & \phn6.1  \nl
G192.8-1.1\dotfill & 4.9$\times 10^{-22}$ & 78             & \phn3.9  \nl
G206.9+2.3\dotfill & 3.8$\times 10^{-22}$ & 60$\times$40   & \phn6.9  \nl
G211.7-1.1\dotfill & 4.6$\times 10^{-22}$ & 70             & \phn4.4  \nl
G261.9+5.5\dotfill & 1.3$\times 10^{-21}$ & 40$\times$30   & \phn5.9  \nl
G272.2-3.2\dotfill & \tablenotemark{c}    & 15             & \tablenotemark{c} \nl
G279.0+1.1\dotfill & 5.0$\times 10^{-22}$ & 95             & \phn3.2  \nl
G284.3-1.8\dotfill & 2.9$\times 10^{-21}$ & 24             & \phn6.0  \nl
G286.5-1.2\dotfill & 1.4$\times 10^{-21}$ & 26$\times$6    & 15.1     \nl
G289.7-0.3\dotfill & 3.7$\times 10^{-21}$ & 18$\times$14   & \phn8.2  \nl
G294.1-0.0\dotfill & \tablenotemark{c}    & 40             & \tablenotemark{c} \nl
G296.1-0.5\dotfill & 1.3$\times 10^{-21}$ & 37$\times$25   & \phn6.6  \nl
G296.8-0.3\dotfill & 4.8$\times 10^{-21}$ & 20$\times$14   & \phn6.9  \nl
G298.6-0.0\dotfill & 7.0$\times 10^{-21}$ & 12$\times$9    & \phn9.5  \nl
G299.2-2.9\dotfill & 3.8$\times 10^{-22}$ & 18$\times$11   & 23.9     \nl
G299.6-0.5\dotfill & 8.9$\times 10^{-22}$ & 13             & 18.1     \nl
G301.4-1.0\dotfill & 3.7$\times 10^{-22}$ & 37$\times$23   & 11.7     \nl
G302.3+0.7\dotfill & 2.6$\times 10^{-21}$ & 17             & \phn8.8  \nl
G308.1-0.7\dotfill & 1.1$\times 10^{-21}$ & 13             & 16.8     \nl
G309.2-0.6\dotfill & 5.9$\times 10^{-21}$ & 15$\times$12   & \phn8.0  \nl
G310.6-0.3\dotfill & 1.2$\times 10^{-20}$ & 8              & 10.0     \nl
G310.8-0.4\dotfill & 6.3$\times 10^{-21}$ & 12             & \phn8.6  \nl
G312.4-0.4\dotfill & 4.7$\times 10^{-21}$ & 38             & \phn3.1  \nl
\tablebreak
G315.4-0.3\dotfill & 3.9$\times 10^{-21}$ & 24$\times$13   & \phn7.2  \nl
G315.9-0.0\dotfill & 3.4$\times 10^{-22}$ & 25$\times$14   & 18.8     \nl
G316.3+0.0\dotfill & 7.4$\times 10^{-20}$ & 29$\times$14   & \phn1.8  \nl
G317.3-0.2\dotfill & 5.8$\times 10^{-21}$ & 11             & \phn9.7  \nl
G318.2+0.1\dotfill & \tablenotemark{c}    & 40$\times$35   & \tablenotemark{c} \nl
G320.6-1.6\dotfill & \tablenotemark{c}    & 60$\times$30   & \tablenotemark{c} \nl
G321.9-1.1\dotfill & \tablenotemark{c}    & 28             & \tablenotemark{c} \nl
G321.9-0.3\dotfill & 2.7$\times 10^{-21}$ & 31             & \phn5.5  \nl
G323.5+0.1\dotfill & 2.7$\times 10^{-21}$ & 15             & 11.4     \nl
G327.4+0.4\dotfill & 1.0$\times 10^{-20}$ & 21             & \phn4.0  \nl
G327.4+1.0\dotfill & 1.8$\times 10^{-21}$ & 14             & 13.9     \nl
G329.7+0.4\dotfill & \tablenotemark{c}    & 40$\times$33   & \tablenotemark{c} \nl
G330.2+1.0\dotfill & 6.2$\times 10^{-21}$ & 11             & \phn9.5  \nl
G332.4+0.1\dotfill & 1.7$\times 10^{-20}$ & 15             & \phn4.5  \nl
G335.2+0.1\dotfill & 5.5$\times 10^{-21}$ & 21             & \phn5.2  \nl
G336.7+0.5\dotfill & 6.5$\times 10^{-21}$ & 14$\times$10   & \phn8.7  \nl
G337.0-0.1\dotfill & 2.8$\times 10^{-20}$ & 13$\times$7    & \phn5.8  \nl
G337.2-0.7\dotfill & 8.4$\times 10^{-21}$ & 6              & 15.3     \nl
G337.3+1.0\dotfill & 1.3$\times 10^{-20}$ & 15$\times$12   & \phn5.6  \nl
G337.8-0.1\dotfill & 5.0$\times 10^{-20}$ & 9$\times$6     & \phn5.9  \nl
G338.1+0.4\dotfill & 2.7$\times 10^{-21}$ & 15             & \phn9.9  \nl
G338.3+0.0\dotfill & 1.6$\times 10^{-20}$ & 8              & \phn8.6  \nl
G340.4+0.4\dotfill & 9.4$\times 10^{-21}$ & 10$\times$7    & 10.5     \nl
G340.6+0.3\dotfill & 3.3$\times 10^{-20}$ & 6              & \phn8.6  \nl
G341.9-0.3\dotfill & 7.2$\times 10^{-21}$ & 7              & 14.0     \nl
G342.0-0.2\dotfill & 4.9$\times 10^{-21}$ & 12$\times$9    & 11.1     \nl
G342.1+0.9\dotfill & 8.4$\times 10^{-22}$ & 10$\times$9    & 25.5     \nl
G343.1-0.7\dotfill & 2.1$\times 10^{-21}$ & 27$\times$21   & \phn6.9  \nl
G345.7-0.2\dotfill & 2.5$\times 10^{-21}$ & 6              & 25.4     \nl
G346.6-0.2\dotfill & 1.9$\times 10^{-20}$ & 8              & \phn8.2  \nl
G348.5-0.0\dotfill & 1.5$\times 10^{-20}$ & 10             & \phn7.2  \nl
G349.2-0.1\dotfill & 3.9$\times 10^{-21}$ & 9$\times$6     & 17.2     \nl
G350.0-1.8\dotfill & 5.2$\times 10^{-21}$ & 30             & \phn3.7  \nl
G351.7+0.8\dotfill & 6.0$\times 10^{-21}$ & 18$\times$14   & \phn6.7  \nl
G351.9-0.9\dotfill & 2.5$\times 10^{-21}$ & 12$\times$9    & 14.7     \nl
G352.7-0.1\dotfill & 1.3$\times 10^{-20}$ & 8$\times$6     & 11.2     \nl
G354.8-0.8\dotfill & 1.2$\times 10^{-21}$ & 19             & 11.1     \nl
G355.6+0.0\dotfill & 9.4$\times 10^{-21}$ & 6$\times$8     & 12.6     \nl
\tablebreak
G355.9-2.5\dotfill & 7.1$\times 10^{-21}$ & 13             & \phn7.6  \nl
G356.3-1.5\dotfill & 1.5$\times 10^{-21}$ & 15$\times$20   & 10.9     \nl
G356.3-0.3\dotfill & 5.9$\times 10^{-21}$ & 7$\times$11    & 12.2     \nl
G357.7+0.3\dotfill & 2.6$\times 10^{-21}$ & 24             & \phn6.2  \nl
G359.0-0.9\dotfill & 6.5$\times 10^{-21}$ & 23             & \phn4.4  \nl
G359.1+0.9\dotfill & 5.7$\times 10^{-21}$ & 11$\times$12   & \phn9.4  \nl
\enddata
\tablenotetext{a}{Surface brightnesses are calculated from fluxes and angular 
diameters given in Green's SNR Catalog (1996a) and Eq. 6 in text.}
\tablenotetext{b}{For roughly circular SNRs, a single diameter is given, 
while for elliptical SNRs, the angular sizes of the major and minor axes 
of the ellipse are given.  Angular diameters are taken from Green's SNR 
Catalog (1996a).}
\tablenotetext{c}{No 1 GHz surface brightness is available because radio 
observations of these SNRs have not been made at enough frequencies, 
preventing an extrapolation of the measured surface brightnesses to 1 GHz. 
Consequently, no distance was derived for these remnants.}
\end{deluxetable}
\clearpage

\section{Selection Effects and the SNR Distribution}

It is of interest to determine the radial distribution of the Galactic 
shell SNRs.  In order to obtain a truer picture of the distribution, the 
problem of selection effects inherent in SNR searches must be addressed.   
Previously, Ilovaisky \& Lequeux (1972) used a derived luminosity 
distribution to correct for selection effects, and based on that, suggested 
a flat SNR distribution out to 8 kpc and a sharp cutoff beyond 10 kpc. Kodaira 
(1974) used empirically determined scaling factors for selection effects and 
obtained a radial distribution peaked at smaller Galactic radii ($\sim5$ 
kpc).  Van den Bergh (1988b) suggested that the fact that the longitudinal 
distribution of SNRs showed a high, uniform concentration between 
$55\arcdeg < l < 345\arcdeg$ could indicate that most of the SNRs in that 
region reside in a nuclear ring with $R=6.5$ kpc.  Leahy \& Xinji (1989) 
assumed circular symmetry around the Galactic Center and near completeness 
of the observations within 2 kpc of the Sun. Using a series of successive 
corrections, they derived an SNR surface density distribution similar to 
Kodaira's and estimated the total number of SNRs in the Galaxy with 
$\Sigma > 3 \times 10^{-22}$ W m$^{-2}$ Hz$^{-1}$ sr$^{-1}$ to be 
$(485 \pm 60)/f_{1}$, where $f_{1}$ is the completeness factor within 2 kpc.

To compensate for observational selection effects we follow a method devised 
by Narayan (1987) for pulsar distributions. This involves calculating a scale 
factor which is the ratio of the number of the predicted shell SNRs in a 
specified volume to the number of observable shell SNRs in that volume. This 
scale factor is determined as a function of Galactic position,
\begin{equation} 
S(r,\phi) = \frac{\int f(\Sigma) d \Sigma}{\int_{obs} f(\Sigma) d \Sigma},
\end{equation}
where $f(\Sigma)$ is the functional form of the distribution of shell SNR 
surface brightnesses and $r$ and $\phi$ are the Galactic radius and azimuth. 
In order to calculate the scale factors, it is necessary to ascertain the 
true SNR surface brightness distribution.  Following Li et al. (1991), we 
assume that the observational surveys for shell SNRs are sufficiently 
complete within 3 kpc of the Sun (this is justified by the scale factors that 
are calculated for that region--see below), and that distribution is taken to 
be representative of $f(\Sigma)$ in the entire Galaxy. The resulting $\Sigma$ 
distribution is shown in Figure~\ref{sb_dist} and can be approximated as a 
Gaussian with a lower cutoff at $\Sigma_{c} = 5 \times 10^{-23}$ W m$^{-2}$ 
Hz$^{-1}$ sr$^{-1}$, corresponding to a remnant age $t \approx 10^{6}$ years.

The Galaxy is divided into bins of 1 kpc $\times$ 10$\arcdeg$ in $r$ and 
$\phi$. A random $r$ and $\phi$ are chosen inside a bin (all points are chosen 
at $z = 0$), and for this point, a corresponding random point is chosen on 
$f(\Sigma)$. One is added to the numerator. Next, based upon its surface 
brightness and angular size, it is determined whether this SNR would be seen 
by the any of the selected radio telescope surveys that were conducted in that 
sky bin. Based on their sensitivities, observing frequencies and amount of sky 
coverage, five telescopes were chosen: Effelsberg 100-m, Parkes 64-m, Dominion 
Royal Astronomy Observatory, Westerbork Synthesis Radio Telescope and Molonglo 
Observatory Synthesis Telescope. The surveys are listed in
Table~\ref{telescopes}. For the single dish telescopes (Effelsberg and Parkes)
the survey sensitivities were calculated following Christiansen \& H\"{o}gbom 
(1985)
\begin{equation}
S_{min} = \frac{2 Q M k (T_{rec} + T_{sky})}{\surd(\Delta \nu t) A_{eff}},
\end{equation}
where $Q$ is the minimum detection significance (here taken to be 5), $M$ is 
a numerical factor dependent upon the type of telescope, $k$ is the Boltzman 
constant, $T_{rec}$ is the receiver noise temperature, $T_{sky}$ is the 
background sky temperature, $\Delta \nu$ is the bandwidth, $t$ is the 
integration time, and $A_{eff} = \eta_{a} \eta_{b} A_{geo}$ is the effective 
area of the telescope, with $\eta_{a}$ and $\eta_{b}$ being the antenna and 
beam efficiencies, respectively. The telescope parameters were taken from the 
references given in Table~\ref{telescopes}. The sky temperature, as a function
of Galactic longitude and latitude, was measured by Haslam et al. (1982) at
408 MHz. Their results were fit by Narayan (1987) to give
\begin{equation}
T_{sky,408}(l,b) = 25 + \frac{275}{\left( 1 + \left( l/42 \right)
^{2} \right) \left( 1 + \left( b/3 \right)^{2} \right) } \mbox{ K}.
\end{equation}
This temperature can be generalized to other frequencies using the frequency
dependence derived by Lawson et al. (1987)
\begin{equation}
T_{sky}(\nu) = T_{sky,408} \left( \frac{408}{\nu} \right)^{2.6} \mbox{ K}.
\end{equation}
For the synthesis telescopes, the sensitivity limit is taken as 5 times the rms
noise in the maps (see Table~\ref{telescopes} for references). If the SNR 
could be detected by at least one of the telescopes surveys that have 
observed that sky bin, 1 is added to the denominator. By repeating for a 
large number of positions on $f(\Sigma)$ an approximate value for $S(r,\phi$) 
is obtained for each bin.

\begin{deluxetable}{lccccc}
\scriptsize
\tablecaption{Telescope surveys used in calculating scale factors 
\label{telescopes}}
\tablehead{
\colhead{Telescope} & \multicolumn{2}{c}{Survey area (deg)} & 
\colhead{Angular res.} & \colhead{Freq.} & \colhead{Ref.} \\
\colhead{} & \colhead{long.} & \colhead{lat.} & \colhead{(arcmin)} & 
\colhead{(GHz)} & \colhead{}} 
\startdata
Effelsberg 100-m\dotfill 
   & $357 \leq l \leq 76$  & $|b| \leq 1.5$ & 4.3 & 2.7   & 1 \nl
   & $76  \leq l \leq 240$ & $|b| \leq 5$   & 4.3 & 2.7   & 2 \nl
   & $357 \leq l \leq 95$  & $|b| \leq 4$   & 9.4 & 1.41  & 3 \nl
Parkes 64-m\dotfill
   & $206 \leq l \leq 49$  & $-2  \leq b \leq 3.5$ & 4.0 & 5.0 & 4 \nl
   & $288 \leq l \leq 307$ & $|b| \leq 2$ & 8.2 & 2.7 & 5 \nl
   & $307 \leq l \leq 330$ & $|b| \leq 2$ & 8.2 & 2.7 & 6 \nl
   & $334 \leq l \leq 345$ & $|b| \leq 2$ & 8.2 & 2.7 & 7 \nl
   & $345 \leq l \leq 5$   & $|b| \leq 2$ & 7.4 & 2.7 & 8 \nl
   & $6   \leq l \leq 26$  & $|b| \leq 2$ & 8.2 & 2.7 & 9 \nl
   & $37  \leq l \leq 47$  & $|b| \leq 2$ & 8.2 & 2.7 & 10 \nl
Dominion Royal Astronomy Observatory\dotfill
   & $81  \leq l \leq 89$  & $-2.5 \leq b \leq 5.5$ & $4.8 \times 3.5$ & 
      .408 & 11 \nl
   & $81  \leq l \leq 89$  & $-2.5 \leq b \leq 5.5$ & $1.4 \times 1.0$ & 
      1.42 & 11 \nl
   & $101 \leq l \leq 109$ & $-2   \leq b \leq 6$   & $3.9 \times 3.4$ & 
      .408 & 12 \nl
   & $101 \leq l \leq 109$ & $-2   \leq b \leq 6$   & $1.2 \times 1.0$ & 
      1.42 & 12 \nl
   & $136 \leq l \leq 144$ & $-2   \leq b \leq 6$   & $3.8 \times 3.4$ & 
      .408 & 13 \nl
   & $136 \leq l \leq 144$ & $|b| \leq 4$           & $1.1 \times 1.0$ & 
      1.42 & 13 \nl
Westerbork Synthesis Radio Telescope\dotfill
   & $44  \leq l \leq 90$  & $|b| \leq 1.5$ & $1 \times 1 \csc \delta$ &
      .327 & 14 \nl
Molonglo Observatory Synthesis Telescope \ldots
   & $335 \leq l \leq 5$   & $|b| \leq 2.5$ & $.73 \times .73 \csc \delta$ & 
      .843 & 15 \nl
   & $240 \leq l \leq 355$ & $|b| \leq 1.5$ & $.73 \times .73 \csc \delta$ & 
      .843 & 16 \nl
\enddata
\noindent
\tablerefs{
(1) Reich et al. 1984; (2) F\"{u}rst et al. 1990; (3) Reich, Reich \& 
F\"{u}rst 1990; (4) Shaver \& Goss 1970; (5) Thomas \& Day 1969a; (6) Day, 
Thomas \& Goss 1969; (7) Thomas \& Day 1969b; (8) Beard, Thomas \& Day 1969; 
(9) Goss \& Day 1969; (10) Day, Warne \& Cooke 1969; (11) Normandeau, Joncas 
\& Green 1992; (12) Joncas \& Higgs 1990; (13) Green 1989a; (14) Taylor, 
Wallace \& Goss 1992; (15) Gray 1994; (16) Whiteoak \& Green 1996}
\end{deluxetable}

This scale factor is then multiplied by the number of SNRs actually observed
to give the corrected number of SNRs in that bin. For bins inside the Solar 
Circle where there are no observed SNRs (hereafter referred to as zero bins),
two different methods were evaluated (and ultimately not used) to assign a 
corrected value to these bins. First, for zero bins which had nonzero bins on 
either side (in the azimuthal direction), the two neighboring bins were 
averaged and that value assigned to the zero bin.  The second method, used 
when two or more consecutive bins had no observed SNRs, utilizes Poisson 
statistics. There is a 50\% Poisson probability of observing zero if the 
expected number is 0.7.  This number (0.7) was divided evenly among the zero 
bins.  However, if this method was applied to the case of an isolated zero 
bin, only 0.7 would be added, whereas a minimum of 1 would have been added 
using the first method. It was eventually decided that neither method would 
be employed, as the two methods did not agree for isolated zero bins, and 
consecutive zero bins may not be expected to have any SNRs observed in them 
(e.\,g. inter-arm regions of the Galaxy). Therefore, no compensation is made 
for a bin where there are no SNRs observed; the corrected number is still 
zero in that bin.

The errors in the scale factors are typically $\sim 10\%$, due mainly to the 
uncertainty in the lower cut-off employed in the surface brightness 
distribution, $f(\Sigma)$, and the telescope sensitivities.
A completeness factor, $f_{c}$, which is essentially the inverse of the 
area-weighted average scale factor, can be estimated to reflect the level of 
observational completeness.  The completeness factor is fairly constant
at $f_{c} \approx 0.98$ until $r = 2.5-3$ kpc, after which the completeness 
factor begins to drop off, justifying our assumption of observational 
completeness out to 3 kpc (see Fig.~\ref{sb_dist}).  Completeness factors, 
$f_{z}$, which are the inverse of the area-weighted average scale factors 
only for the zero bins, can also be estimated to reflect the level 
incompleteness due to zero bins.  For the entire near half of the Galaxy (see 
Fig.~\ref{distrib}), $f_{z} \sim 0.85$ for $\Sigma > 5 \times 10^{-23}$ W 
m$^{-2}$ Hz$^{-1}$ sr$^{-1}$.  For the region within 3 kpc of the Sun, 
$f_{z} \approx 0.99$.  

The bins, corrected for selection effects, are combined into three broad 
regions in azimuth.  These regions are shown in Fig.~\ref{distrib}. The 
apparent absence of SNRs in the outer portion of region 3 is probably real, 
due to the outer spiral arms ending before reaching this region. The bins in 
each region are combined further in azimuth to obtain a single radial profile 
for that region and converted to surface densities.  When 1 kpc radial bins 
were used, peaks in the radial distribution show a rough correspondence to 
the positions of the spiral arms.  In order to get an estimate of the errors 
in the shell SNR surface densities due to the uncertainty in the $\Sigma-D$ 
relation, a Monte Carlo simulation was used, taking the fractional errors of 
Fig~\ref{frac_err} into account, in order to obtain bootstrapped errors.  The 
errors in the densities are typically $\sim10-20\%$ in the mid-regions of the 
Galaxy ($r \approx 4-12$ kpc) and considerably higher (up to 75\%) in the 
regions of low statistics ($r \leq 4 $ kpc, $r \geq 12$ kpc).  The scale 
factors cannot compensate for bins where no SNRs are observed, and therefore 
observational incompleteness is still a problem for regions 1 and 3, where 
the completeness factors are lower. The scaled total number of shell SNRs in 
region 2 is $(56 \pm 4)/f_{z}$, where the error on the number of SNRs 
represents the uncertainty in the $\Sigma-D$ relation and $f_{z}$ represents 
the incompleteness due to the lack of selection effects compensation for the 
zero bins. For region 2, $f_{z} \approx 0.96$.  If region 2 is considered 
representative of the entire Galaxy, then the total number of shell remnants 
for $r \leq 16$ kpc and $\Sigma > 5 \times 10^{-23}$ W m$^{-2}$ Hz$^{-1}$ 
sr$^{-1}$ is estimated to be $336/f_{z}$.  The Monte Carlo simulation shows 
that this estimate is not very sensitive to the uncertainty in the 
$\Sigma-D$ relation.

\begin{figure}[tb]
\begin{center}
\epsfxsize=6.5in
\epsfbox{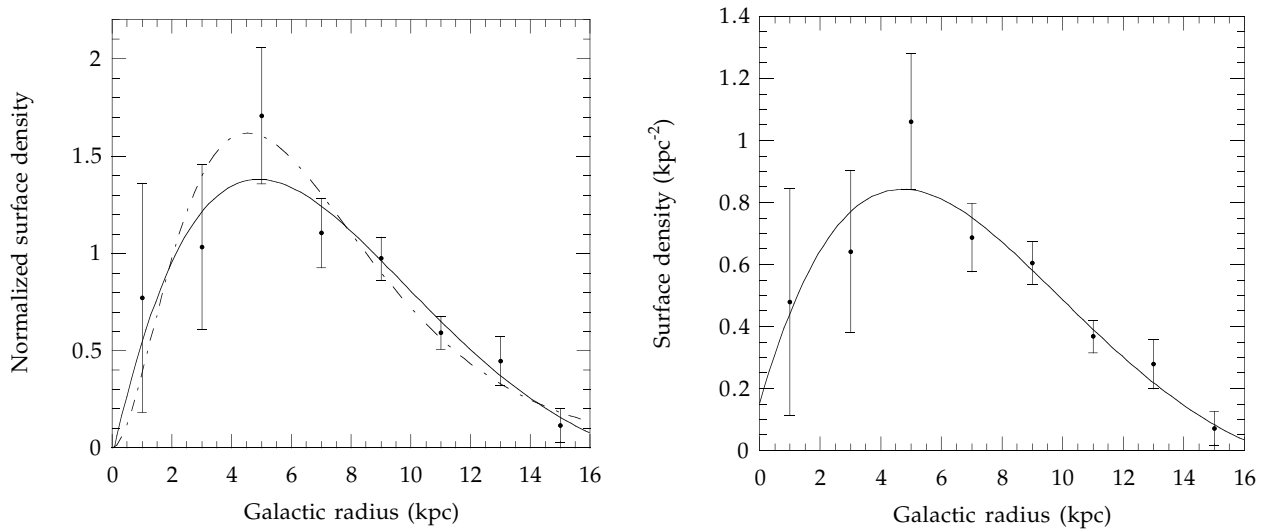}
\caption{The SNR density radial distribution for 
Region 2 using the new distances and compensation for selection effects. (a) 
The distribution derived in this work (solid line and data points) and that 
of Kodaira (1974) (dashed line), both normalized to the density at the radius 
of the Solar Circle.  (b) The unnormalized data points and the fit to Eq. 15.  
\label{radial}}
\end{center}
\end{figure}

A weighted fit of the shell SNR surface density distribution in region 2, 
normalized to the surface density at the Solar Circle, was 
performed using the functional form employed by Stecker \& Jones (1977),
\begin{equation}
f(r) = \left( \frac{r}{r_{\odot}} \right) ^\alpha \rm{exp} \left( -\beta
\frac{r - r_{\odot}}{r_{\odot}} \right),
\end{equation}
where $r_{\odot} = 8.5$ kpc is the Sun-Galactic Center distance.  We find that 
$\alpha = 2.00\pm0.67$ and $\beta = 3.53\pm0.77$; the radial scale length 
of the distribution is $\approx7.0$ kpc.  The shape of the distribution is 
similar to that obtained by Kodiara (1974). The two distributions are shown in 
Figure~\ref{radial}a.

Equation 14 implies that the surface density is zero at $r=0$.  However, our 
data suggest that the surface density is not zero near the Galactic Center.  
Therefore, we have used the following functional form to obtain a weighted 
fit to the unnormalized surface density distribution:
\begin{equation}
f(r) = A \sin \left( \frac{\pi r}{r_{0}} + \theta_{0} \right) e^{-\beta r}
\end{equation}
where $A=1.96\pm1.38$ kpc$^{-2}$, $r_{0}=17.2\pm1.9$ kpc, 
$\theta_{0}=0.08\pm0.33$ and $\beta=0.13\pm0.08$ kpc$^{-1}$.  This fit is 
valid for $r < r_{o} (1-\theta_{o}/\pi)$, i.e. 16.8 kpc; $f(r)=0$ beyond that.
The data and fit are shown in Figure~\ref{radial}b.

The scale length of 7.0 kpc is consistent with that determined by previous 
studies.  Green (1996b) used a simple model with SNRs distributed as a 
gaussian in Galactic radius and compared the resulting longitudinal 
distribution to the observed SNR longitudinal distribution, obtaining a scale 
length of $\approx 7.0$ kpc.  However, no attempt was made to compensate for 
selection effects other than to use a $\Sigma$-limited sample.  Li et al. 
(1991) used a more sophisticated model distributing SNRs in an exponential 
disk as well as in spiral arms.  They incorporated a 1/$d^{2}$ selection 
bias, assuming completeness out to $d = 3$ kpc.  They then compared the 
longitudinal distribution given by the model with the observed SNR 
longitudinal distribution, obtaining a scale length of $\approx 5-9$ kpc, 
depending on the model parameters.  As Li et al. point out, the scale length 
of the Galactic stellar disk is $\sim 5$ kpc, suggesting that the SNR scale 
length, as derived in this work and by Green (1996b) and Li et al. (1991),  
would indicate that the SNR distribution is not associated with the stellar 
disk population.

\section{Conclusion}
The catalogue of known SNRs has continued to grow in size.  The number of 
SNRs with reasonably determined distances has also increased.  However, most 
distances given in the literature were calculated using older rotation curves. 
We have recalculated the distances, where necessary, using a modern rotation 
curve, and used the updated distances to derive a new $\Sigma-D$ relation for 
shell SNRs. This $\Sigma-D$ relation, using a sample of 36 shell SNRs (37 
including Cas A), yields a slope of $-2.38$ excluding Cas A and $-2.64$ with 
Cas A.  When the 41 shell SNRs in the LMC and SMC are added to the sample, 
the slope is $-2.41$ with a smaller error.  Using the $\Sigma-D$ relation to 
estimate distances to individual remnants is viable only with the assumptions 
that all shell SNRs have the same radio luminosity dependence on linear 
diameter, the same supernova explosion mechanism and energy, and are evolving 
into identical environments.  We find that, on average, the error in the 
distance estimation to an individual SNR to be $\sim40\%$ when using our 
$\Sigma-D$ relation. However, the error in deriving ensemble characteristics 
of SNRs such as the SNR surface density can be lower ($\sim 10-20\%$ for the 
mid-galactic region). We attempt to compensate for observational selection 
effects inherent in SNR searches by employing a scaling method based on the 
sensitivity, angular resolution, and sky coverage of actual radio surveys. 
Utilizing the updated distances, the new $\Sigma-D$ relation and the scale 
factors, the shell SNR surface density radial distribution was derived.  The 
distribution peaks at $\sim 5$ kpc and has a scale length of $\sim 7.0$ kpc.

\acknowledgments
We would like to thank Robert Hurt for useful discussion on radio telescope 
surveys, as well as anonymous referees for helpful suggestions. We also 
would like to thank the staff of the U. C. Riverside IGPP for their support.  
This work was supported in part by NASA through the grant NAGW-1996.

\end{document}